# Design of Small Multi-band Full-screen Smartwatch Antenna for IoT applications

Bing Xiao, Hang Wong, Di Wu, and Kwan L. Yeung

*Abstract*—**Smartwatch is a potential candidate for the Internet of Things (IoT) hub. However, the performance of smartwatch antennas is severely restricted by the smartwatch structure, especially when the antennas are designed by traditional methods. For adapting smartwatches to the role of IoT hub, a novel method of designing multi-band smartwatch antenna is presented in this paper, aiming at increasing the number of frequency bands, omni-directivity, and structural suitability. Firstly, the fundamental structure (including the full screen and the system PCB) of the smartwatch is analyzed as a whole by characteristic mode analysis (CMA). Thus, abundant resources of characteristic modes are introduced. The fundamental structure is then modified as the radiator of a multi-band antenna. Then, a non-radiating capacitive coupling element (CCE) excites the desired four $0.5\lambda$ modes from this structure. This method could fully utilize the intrinsic modes of the smartwatch structure itself, thus exhibits multiple advantages: significantly small size, smaller ground, omni-directional radiation, and fitting to the full-screen smartwatch structure.**

## I. INTRODUCTION

Smartwatches are more and more prevalent as convenient wearable devices, becoming a potential candidate for the hub of the Internet of Things (IoT) [1], [2] and Body Area Network (BAN) [3]. Numerous studies have been conducted on their applications in IoT. In [4], a contact tracing framework is proposed for logging users' co-existence by received signal strength indicators (RSSI) provided by carried IoT devices, such as smartwatches and smartphones. The interacting pair, the estimated distance, and the overlapping time duration were recorded. This framework is beneficial especially during pandemic outbreaks. In [5], a system for gender recognition was proposed by monitoring the walking patterns. Accelerometer and gyroscope signals produced by smartwatches were recorded at a fixed interval. An artificial neural network was utilized to classify the gender after the raw data were preprocessed. In [6], a prototype based on smartwatches was proposed to help students with disabilities for daily life tasks inside a building, including environmental monitors and in-building device controlling. In [7], a system was proposed employing smartwatches and smartphones to detect the driver's dangerous behaviors in a real-time manner. In [8], a smartwatch-based IoT system was proposed to monitor multiple hospitalized patients. In case of alert notifications, urgent medical interventions could take effect immediately. In [9], a low-cost IoT based system monitoring acoustic, olfactory, visual, and thermal comfort levels was designed. It was provided by a smartwatch with computing, controlling, connecting modules, and various ambient sensors. In [10], a survey and framework based on context computation, edge analytics, and computation off-loading achieved by smartwatches was presented, aiming at overcoming the weakness of the short battery life of smartwatches. In [11], the recognition accuracy of activities, such as basketball, was improved by combining the data from an accelerometer, a gyroscope, and an audio sensor produced by the *Samsung Gear S3* smartwatch. In [12], pen-holding gesture recognition was achieved by utilizing the sensors of smartwatches. In [13], a pointing approach to interact with smart devices by the inertial sensors in smartwatches was proposed. An ultra-wide band (UWB) transceiver for identifying the selected device from the ranging measurement is utilized.

In order to achieve the above-mentioned ever-increasing various IoT functions, qualified radio frequency communication systems for smartwatches are necessary. At present, 2.4-GHz Wireless Local Area Network (WLAN) and Bluetooth are the most common wireless communication protocols adopted in commercially available smartwatch products. Radio frequency identification (RFID), Zigbee [14] could also utilize this 2.4-GHz industrial, scientific, and medical (ISM) band. However, an IoT hub has to be compatible with multiple frequency bands in multiple communication protocols for connecting various IoT devices. The single frequency band is far from satisfying this requirement. For example, the emerging 5G communication supports the function of massive machine-type communication (mMTC) [15], connecting to massive IoT terminals by intermittently transmitting small amounts of traffic. Moreover, the global navigation satellite system (GNSS) is essential for accurate positioning.

However, it is challenging designing multi-band antennas with good performance on smartwatches. Firstly, the size of a smartwatch is very small. The antennas could only occupy a very limited area and volume, because many other components are competing for the space. Secondly, the recent trend of full-screen smartwatches further squeezes and significantly affects the smartwatch antennas [16], [17]. However, at the same time,

This work was supported in part by the Research Grants Council of the Hong Kong SAR, China with a TRS project (Project No. T42-103/16-N), a CRF project (Project No. CityU C1020-19E), and the CityU Strategic Research Grant (Project No. SRG-Fd 7005227).

Bing Xiao and L. K. Yeung are with the Department of Electrical and Electronic Engineering, The University of Hong Kong, Hong Kong.

H. Wong is with The State Key Laboratory of Terahertz and Millimeter Waves, and Shenzhen Research Institute, Department of Electrical Engineering, City University of Hong Kong, Hong Kong (e-mail: hang.wong@cityu.edu.hk).

Di Wu is with College of Electronic and Information Engineering, Shenzhen University, Shenzhen, China.

most present literature could not take the full screen, which contains conducting layers, into consideration. Thirdly, most smartwatches are too small to be equipped with a beam-steering antenna array, especially for low GHz bands. Thus, an omnidirectional radiation pattern is important for smartwatch antennas, because they could utilize a single antenna to keep steadily transmitting and receiving signals at any angle. However, smartwatch antennas in the present literature rely on higher-order modes to design multi-band antennas, inevitably introducing more nulls in the radiation patterns [18]–[20].

In the present literature, the majority of smartwatch antennas concentrated on 2.4-GHz WLAN/Bluetooth band. In [21], a shorted monopole antenna was proposed for the 2.4-GHz frequency band. It is mounted conformally along the side surfaces as an internal antenna. In [22], an open slot Bluetooth antenna on the metal frame of the smartwatch is designed, excited by a 50 Ω microstrip feedline. In [23], a pair of degenerated modes on the metal frame was applied in designing a high-isolation multiple-input and multiple-output (MIMO) antenna for the 2.4-GHz band. In [24], MIMO slot inverted-F antennas are designed by extending the two sides of the ground. In [25], the antenna operated in one-wavelength resonant mode on the metal frame to achieve the 2.4-GHz band. In [26], a cavity-backed annular slot antenna was proposed for the 2.4-GHz band. The specific absorption rate (SAR) and the effect of human tissue are significantly reduced. In [27], the smartwatch antenna was incorporated with a layer of electromagnetic band-gap (EBG) structure to reduce SAR and backward radiation towards the body. In [28], the antenna made by the watch belt, instead of on the watch body, was introduced. Thus, it could have enough length and cover multiple frequency bands, including LTE bands. However, the antenna with the watch belt suffers from higher loss from the large-area contiguous human skin. In [29], a miniature high impedance surface (HIS) was proposed for the 2.4-GHz smartwatch antenna to reduce the effect of the human tissue.

In addition to the above-mentioned single-band antennas, there are also multi-band antennas. In [18], a broken metal rim served as the major radiator. By a feeding point and a shorting strip placed at proper locations, the antenna could cover four frequency bands. However, the indispensable 2.4-GHz band is not included. In [19], the proposed antenna utilized different loop modes on the metal frame to cover GPS, 2.4-GHz, and 5-GHz WLAN bands. In [20], an unbroken metal rim was utilized to design a triple-band antenna. It used different loop modes to cover 2.4-GHz, 3.5-GHz, and 4.9-GHz bands. In [30], an unsymmetrical slot antenna formed by the metal frame and the internal copper cylinder is designed for dual-band operation. In [31], an optically transparent rectangular antenna mounted on the smartwatch screen is achieved by applying indium tin oxide (ITO). It could cover both 2.4-GHz and 5-GHz WLAN bands.

We know that mobile devices have both antenna-element wave modes and PCB (chassis) wave modes [32]. Both of them can be utilized to design antennas. In the existing literature, some only used antenna-element wave modes, such as PIFAs, monopole antennas [33], metal frame antennas [18]–[20], metal frames with PIFAs [34]. Some only used PCB wave modes [35]. Others used both of them but analyzed and designed them separately [17], [32].

In this paper, we expand the concept of smartwatch antenna design to the overall fundamental structure, especially including the full screen, in addition to the PCB. By small modifications with a shielding zone and a short stub, the overall fundamental structure of the smartwatch serves as a radiator. A short non-radiating capacitive coupling element (CCE) excites the desired multiple modes. This novel method could fully utilize the intrinsic modes of the smartwatch structure itself, thus exhibit multiple advantages. The designed antenna covers up to four bands, including GPS L1 1575-MHz band, WLAN/Bluetooth 2.4-GHz band, 3.5-GHz 5G band (N78), and 4.9-GHz 5G band (N79). Only two short strips are utilized: a subsidiary stub and a CCE, taking only 36% of the perimeter in total. The antenna is applicable to a full-screen smartwatch design, which is widespread at present. This antenna is applicable to a smaller ground instead of a full-size PCB. Moreover, due to the abundant resources of modes, we could only select the half-wavelength dipole modes. Thus, this antenna exhibits omni-directional radiation patterns at all operating frequency bands.

For this paper, in Section II, the process of designing a smartwatch antenna with the theory of characteristic modes (TCM) is presented. In Part A, the structure of the smartwatch is analyzed, and the reason is given for applying TCM. Then TCM is briefly reviewed in Part B. The shielding zone of the flexible printed circuit (FPC) for GPS L1 band is introduced in Part C. The additional stub to the PCB for 2.4 and 3.5 GHz bands is presented in Part D. The transverse mode for 4.9 GHz band is in Part E. Three critical parameters are analyzed in Part F. At last, the method of excitation is presented in Part G. In Section III, the prototype and measured results of the designed antenna is presented.

## II. CHARACTERISTIC MODE ANALYSIS

### A. Structure of Smartwatch

The external size of the smartwatch model is 44 mm × 38 mm × 10.7 mm, the same as the *44 mm Apple Watch Series 4* [36]. In the smartwatch, the fundamental metallic structures are the screen assembly and the PCB. Other structures could be seen as attached to them.

At present, most of the existing literature could not take the screen into consideration. However, in recent years, one obvious trend for smartwatches and other smart devices is the growth of the screen-to-body ratio. The concept of full screen is prevalent. The metal-contained large screen assembly will significantly degrade the radiation performance of the very close antennas. For example, the bandwidth of the antenna will be narrowed, as presented in [16].

The screen assembly is on the front side of the smartwatch. It is used for graphics display and human-machine interaction. For the full-screen smartwatch, the screen covers the whole front side of the smartwatch body. In the proposed design, the gap width between the screen and the sides of the watch is only 1 mm, so the size of the screen is 42 mm × 36 mm. It means the screen-to-body ratio reaches 90.4%. The structure of the screens of smartphones has been analyzed in detail in [16]. The structure of the smartwatch's screen assembly is the same. Generally it contains, from outside to inside, the glass cover, the touch panel, the liquid crystal display (LCD), the backlight, and the metal cover plate. Especially, the metal cover plate is at the bottom of the screen assembly, for enhancing the structural strength of the screen assembly, and separating the screen assembly from other interior components for shielding the electromagnetic interference between them. These layers are combined closely. For the theoretical study of antennas, it is impractical to include such a complex structure in the electromagnetic simulation and later prototyping process. Thus, this multi-layer screen is simplified and modeled as a layer of the perfect electric conductor (PEC). Fig. 1 shows the comparison of the simulated results of reflection coefficient and efficiency between a single layer of PEC and a smartphone screen (LCD type). The simulation is conducted by CST Studio Suite. The emulated screen is 1 mm thick in total, composed of five representative layers of a real LCD screen, as indicated in [16]. They are Corning glass, polymethyl methacrylate (PMMA), liquid crystal, steel, and graphite (from outer to inner layer). From the simulated results, we can see that the PEC model of the screen could well emulate the 5-layer electromagnetic model, which represents the real multi-layer complex-structured screen, especially in the aspect of the reflection coefficient. However, the total efficiency of the antenna is lower in the screen model than in the PEC model in the whole spectrum. That is because the lossy materials in the screen absorb a small part of the radiated/received power from/to the antenna.

| Layers of simulated screen | Relative permittivity | Loss tangent | Electrical conductivity (S/m) | Thickness (mm) |
|---|---|---|---|---|
| Corning® 7059 glass | 5. 84 [37] | 0.0036 [38] | - | 0.2 |
| PMMA | 2.58 | 0.009 | - | 0.2 |
| Liquid crystal | 3. 2.5 [39]* | 0.03, 0.13 [39] * | - | 0.2 |
| Steel 1008 | - | - | 7.69e+6 | 0.2 |
| Graphite | - | - | 1e+5 | 0.2 |

* εr and εL @ 3GHz

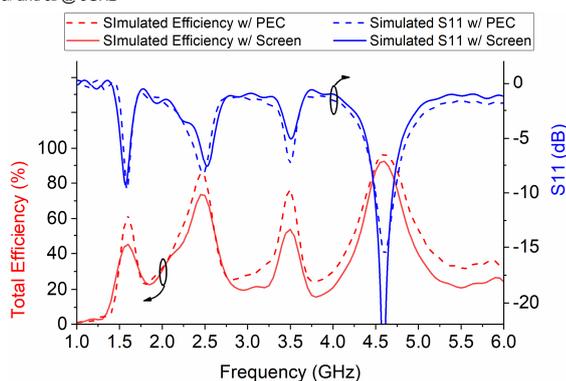

Fig. 1. Comparison of simulated reflection coefficient and total efficiency with PEC and screen respectively

The system PCB is below and parallel to the screen. The size of the PCB is set as 34 mm × 28 mm, only 56.9% of the area of the smartwatch body. The screen and the PCB are electrically connected by a piece of flexible printed circuit (FPC). FPC is an essential part of the screen. It is used for transmitting and receiving signals between the screen and the system PCB. Same with the condition of practical products, the FPC is also included in the antenna model. It is simplified and modeled as a PEC narrow strip connecting the screen and the PCB [16]. The spacing between the screen and the PCB is only 4 mm. Because the total height of the smartwatch is 10.7 mm, the PCB is in the middle of the smartwatch body. Below the PCB, there could be other components in practical products, such as the heart rate monitor [40].

Because of the above-mentioned intrinsic structure of the smartwatch, the antenna's *ground* of the system PCB turns to a *ground* of H-shaped connected screen and PCB. In this situation, it is not suitable for us to directly apply the classical analytical methods, which are based on a single-plate ground [33]. Instead, we resort to TCM.

### B. Introduction to theory of characteristic modes

The theory of characteristic mode (TCM) could provide characteristic modes of an arbitrarily shaped structure, without considering the feeding arrangement. It brings physical insight into the radiating phenomena of the antenna. The characteristic modes are obtained by solving an eigenvalue equation derived from the method of moments (MoM) [41], [42]:

$$X(\mathbf{J}_n) = \lambda_n R(\mathbf{J}_n) \qquad (1)$$

Where $X$ and $R$ are the imaginary and real parts of an impedance matrix $Z$. $\mathbf{J}_n$ is the eigencurrent of the $n$-th mode on the structure. $\lambda_n$ is the corresponding eigenvalue. The current on a conductive surface can be expressed as a linear superposition of the normalized eigencurrents. Besides, the corresponding characteristic angle of the $n$-th mode can be computed from [43]:

$$\theta_n = 180° - \arctan \lambda_n \qquad (2)$$

At a given frequency, we can determine whether or not the characteristic mode is at resonance by looking at its characteristic angle. When the characteristic angle $\theta_n$ is near $180°$, the mode is at resonance. Because the curves of characteristic angles are mostly monotonically crossing the $180°$ line, the resonant frequencies of each mode could be clearly observed, especially when we conduct parameter sweeps. So in this paper, we select the characteristic angle as the primary indicator for the characteristic mode analysis. The following characteristic mode results are all simulated by the CMA solver of CST Studio Suite. It is noteworthy that because of the immature algorithm of CST, the mode tracking may have errors when spreading in a wide frequency range [44]. However, because we only observe the eigencurrent distribution and frequency deviation of each mode at exactly the resonant frequency of the concerned mode (when its characteristic angle $\theta_n = 180°$), our results and conclusions are not affected by the inaccurate mode tracking. The 3D electromagnetic simulations are all conducted by the time domain solver of CST Studio Suite.

In the following parts, the TCM method of modifying the smartwatch structure for designing an antenna is presented step by step, from lower to higher frequency bands, as shown in Fig. 2. For a clear observation of the inner structure, the screen on the front side of the smartwatch is shown in semi-transparent. Firstly, the shielding zone is designed for the lowest GPS L1 frequency band. Secondly, an additional stub extended from the PCB is designed for the 2.4-GHz band. Thirdly, the stub is extended in the opposite direction for the 3.5-GHz band.

Fourthly, the longer edge of the PCB is utilized for the 4.9-GHz band. Lastly, a short CCE is mounted to excite the desired four modes.

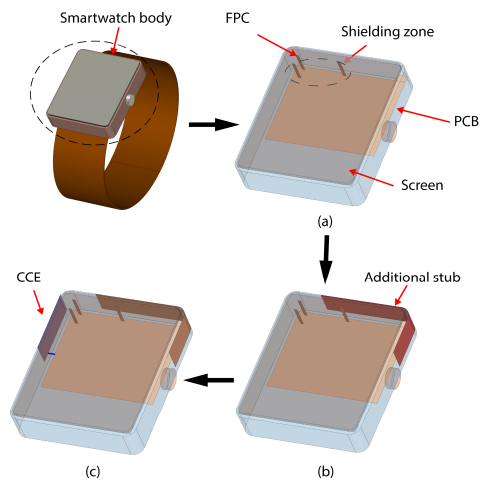

Fig. 2. Step-by-step design of the smartwatch antenna by adding (a)Shielding zone, (b) Additional stub, (c) CCE for excitation

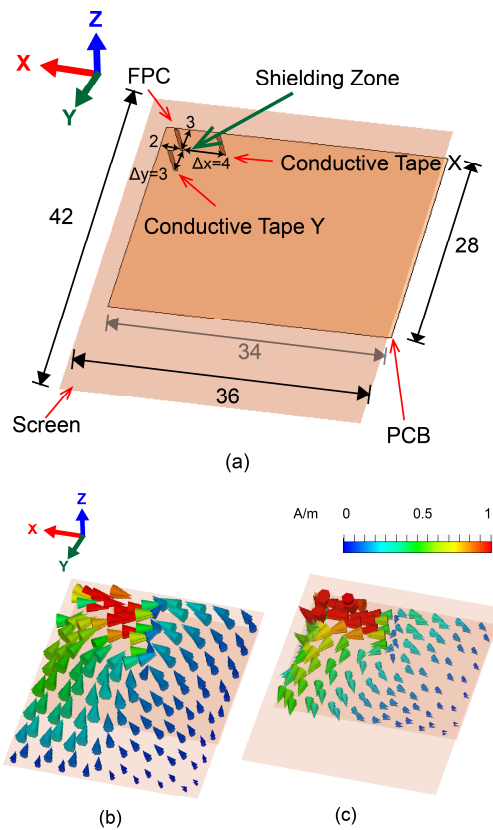

Fig. 3. Schematic of (a) smartwatch's main architecture and normalized eigencurrent of Mode 1 on (b) screen and (c) structures other than screen

### C. Shielding zone of FPC for GPS L1 band

As shown in Fig. 3(a), for the structure of the screen and the PCB connected by the FPC, the fundamental eigenmode could resonate at around the desired 1.575 GHz because it has enough electrical length. However, the exact control of the resonating frequency is very difficult in the practical developing processes of smartwatches. Firstly, the exact size of the PCB is usually decided by the circuit engineers instead of the antenna engineers. It mainly depends on component placement and circuit routing on the PCB. Secondly, the size of the screen is usually predefined, sorted from the limited types of sizes provided by the screen manufactures and recognized by the market. At last, the exact location of the FPC is also not decided by the antenna designer, but the screen and circuit engineers. Because the FPC is an essential part of the screen assembly, thus its location mainly depends on the implementation of the functions of display and touch control. Thus, the resonating frequency is fixed in a way before the antenna design. Cutting slot is a method proposed to change the resonating frequency of the PCB [45]. However, the slot on the whole piece of PCB will interrupt the wire routing of every layer of the PCB, generally eight, ten, or even twelve layers, increasing the difficulty of wire routing drastically. So, this method is seldom used in practical products.

To solve this problem, we make a shielding zone of the FPC by conductive tapes. As shown in Fig. 3(a), the conductive tapes connect the screen and the PCB, around the FPC. Thus, the FPC is only required to stay in the shielding zone instead of at an exact location. It is certain that the more conductive tapes around the FPC, the more effective the shielding zone will be. Here, we try to use just two conductive tapes in x- and y-directions respectively, namely *Conductive Tape X* (*CTX*) and *Conductive Tape Y* (*CTY*). This measure could balance effectiveness and structure simplicity. Then we try to tune the resonating frequency of this fundamental structure mode (Mode 1) by changing the locations of these two tapes. The simulated normalized eigencurrent distribution is shown in Fig. 3(b) and (c). The eigencurrent goes along the screen, mostly goes through the conductive tapes, then goes backward on the PCB. The total electrical length is about a half wavelength of 1575 MHz. The nearer the shielding zone to the top-left corner of the watch, the longer the current path, the lower the resonant frequency.

As shown in Fig. 3(a), the origin of the coordinate system is at the top-right corner of the PCB. $\Delta x$ is the difference in x-axis between the FPC and *CTX* ($\Delta x = X_{FPC} - X_{CTX}$). $\Delta y$ is the difference in y-axis between *CTY* and the FPC. ($\Delta y = Y_{CTY} - Y_{FPC}$). When $\Delta x$ and $\Delta y$ are all positive, the FPC locates in the shielding zone. When the shielding zone is fixed, the variation of the characteristic angle of Mode 1 is shown in Fig. 4. The FPC moves along x- and y-direction. As can be seen, the resonating frequency is significantly less sensitive to the FPC's location when it is in the shielding zone, compared with outside the shielding zone. It shows that the two-conductive-tape shielding zone is effective in this model. However, on the other hand, the bandwidth of this mode will be narrowed by these two narrow strips. Because for the $0.5\lambda$ dipole mode, the wider the resonant structure, the broader the fractional bandwidth.

Furthermore, Fig. 5 shows the radiation pattern of Mode 1. It is a typical doughnut-shaped pattern resulted from a $0.5\lambda$ dipole. When the smartwatch is at the natural angle by which people are watching it, the doughnut is perpendicular to the earth. This characteristic makes this mode suitable for the GPS application.

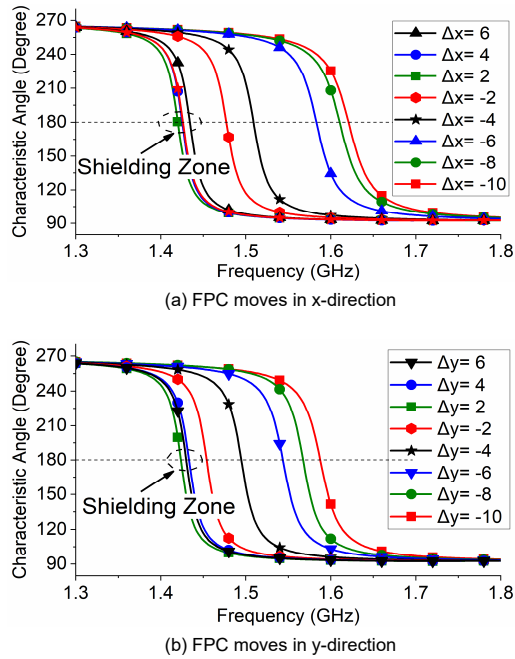

(a) FPC moves in x-direction

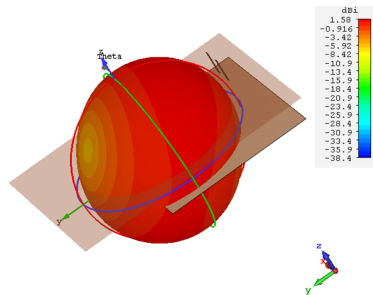

(b) FPC moves in y-direction

Fig. 4. Variation of characteristic angles of Mode 1 when the FPC moves in x- and y-direction (unit: mm)

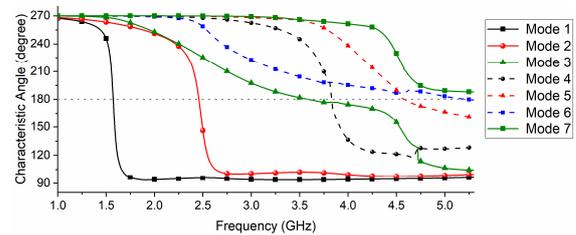

Fig. 5. Radiation pattern of Characteristic Mode 1

In a word, by taking this measure, we could tune the resonating frequency of Mode 1 to the GPS frequency band only by moving *CTX* and *CTY*, thus reduce the influence of the FPC's location significantly.

### D.  Additional stub for 2.4-GHz and 3.5-GHz bands

For an electrical length of half wavelength of 2450 MHz, we add a stub to extend the PCB, as shown in Fig. 6(a). Then the modified smartwatch structure could produce a 2450 MHz Mode (Mode 2). The normalized eigencurrent distribution of this mode is shown in Fig. 6(b) and (c). Different from Mode 1, its eigencurrent on the screen is quite weak. It generally goes along the PCB and then the stub. Thus, the longer the stub, the lower the resonant frequency. The total electrical length should be about a half wavelength of 2450 MHz.

In order to cover the 3500 MHz frequency band, the stub is extended in the opposite direction along the smartwatch perimeter, as shown in Fig. 7. For this mode (Mode 4), the eigencurrent concentrates on the L-shape stub. Thus, the longer the extended stub in the positive y-direction, the lower the resonant frequency. The total electrical length of the whole L-shape stub should be about a half wavelength of 3500 MHz.

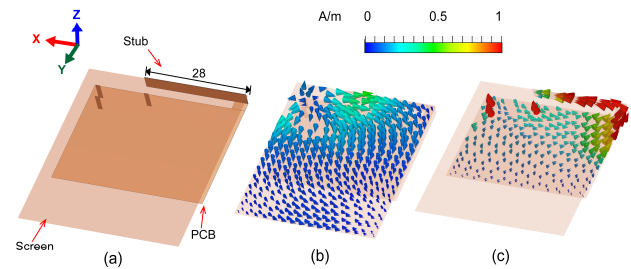

Fig. 6. Schematic of (a) smartwatch's structure with a short stub added and its normalized eigencurrent of Mode 2 on (b) screen and (c) structures other than the screen

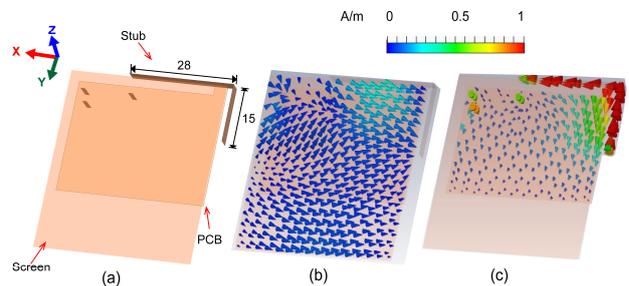

Fig. 7. Schematic of (a) smartwatch's structure with a stub extended in the opposite direction and its normalized eigencurrent of Mode 4 on (b) screen and (c) structures other than screen

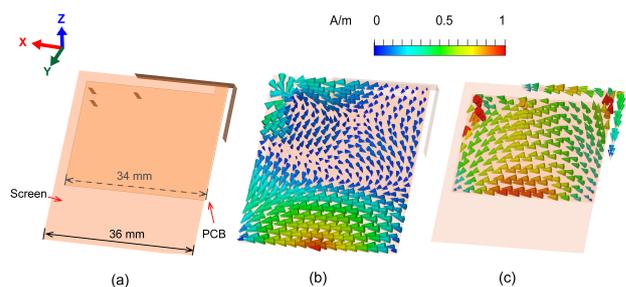

Fig. 8. Characteristic modes of the modified structure

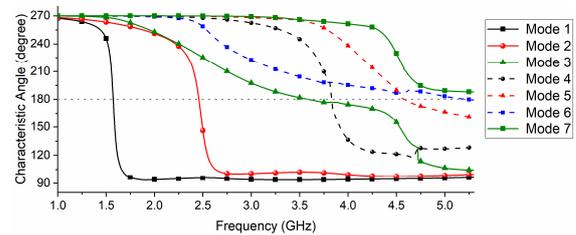

Fig. 9. Schematic of (a) smartwatch's structure and normalized eigencurrent of Mode 7 on (b) screen and (c) structures other than screen

### E.  Transverse mode for 4.9-GHz band

Until now, for the modified watch structure, the lowest several characteristic modes are shown in Fig. 8. Among them, Modes 1-3 have been utilized. As already exhibited in a lot of literature, the resource of characteristic modes of any structure is always getting richer as the frequency increases.

We can see the length of the longer edge of the PCB is 34 mm, and the width of the screen is 36 mm, as shown in Fig. 9(a). They are close to the half wavelength of 4900 MHz. The eigencurrent of Mode 7 is mainly on the above-mentioned two edges with $0.5\lambda$ distribution in the same phase, shown in Fig. 9(b) and (c). In addition, due to the large size of the screen and the PCB, this mode could have wider bandwidth. Thus, this mode will also be used. It is noteworthy that, as mentioned previously, the sizes of the PCB and the screen are usually predefined. The resonating frequency of this mode will not definitely be 4900 MHz. Thus, instead, we will tune it by the impedance matching network later.

Apart from the above-mentioned Modes 1-3 and 7, other modes could also be utilized, tuned, and properly excited if even more frequency bands are required.

### F. Parameter sweeps on some critical dimensions

Moreover, we conduct parameter sweeps on three critical variable dimensions: the length of the upper stub $L_1$, the length of the right stub $L_2$, and the width of the PCB $YY$, indicated in Fig. 11. The results are shown in Fig. 10. When $L_1$ decreases from 30 mm to 24 mm, the resonant frequencies of Mode 2 and Mode 4 increase. The reason is as analyzed in Section II(C) and Section II(D): the eigencurrents of these two modes all flow through the upper stub with a length of $L_1$. It is noteworthy that Mode 1 also has a slight frequency deviation even if it is not designed by purpose. This phenomenon indicates that apart from the H-shaped main architecture of the smartwatch as Fig. 3, there is also a small part of eigencurrent of Mode 1 flow through the upper arm because the electrical length of this current path is close to the designed one.

In Fig. 10(b), when $L_2$ decreases from 20 mm to 14 mm, the resonant frequency of Mode 4 increases, while the frequencies of other modes keep unchanged. That is because only the eigencurrent of Mode 4 flows through the right stub.

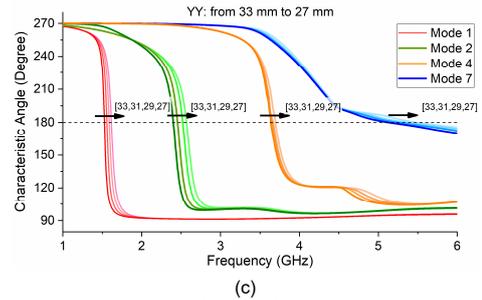

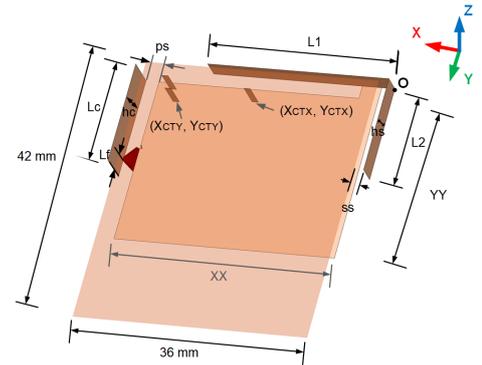

Fig. 10. Parameter sweeps on (a) length of the upper stub $L_1$, (b) length of the right stub $L_2$, and (c) width of PCB $YY$

Fig. 11. The structure and dimensions of the proposed antenna

In Fig. 10(c), when $YY$ decreases from 33 mm to 27 mm, the resonant frequencies of Mode 1 and Mode 2 increase, because the eigencurrents of these two modes all flow along the Y-axis. When $YY$, the width of the PCB decreases, the electrical length is shortened. Mode 4 varies slightly because its eigencurrent mainly concentrates on the upper and right stubs. Mode 7 changes slightly because its eigencurrent is mainly along the X-axis.

### G. Excitation

The coupling elements are responsible for exciting the desired characteristic modes without contributing themselves too much to the radiation. Due to the close vicinity between the radiator (here is the modified overall fundamental structure of the smartwatch) and the coupler, the field on the radiator could be excited by the coupler. Coupling elements are frequently used in the TCM method. There are two types of them: capacitive coupling element (CCE) and inductive coupling element (ICE). It is best for CCE to locate at the maxima of the electric eigenfield (corresponding to the minima of the eigencurrent when there is a standing wave). ICE is just the opposite [46], [47].

Both of them could be selected for exciting. However, a research was conducted in [46] to compare the purity of excitation between CCE and ICE. It showed that CCE is usually less pure in exciting a single mode compared to ICE. The reason is that for different modes, CCEs tend to locate at the corners, where many modes possess maxima of the electric eigenfield there. On the opposite side, this phenomenon also means that CCEs are easier to excite multiple eigenmodes simultaneously, which is just required here. So we select CCEs for exciting.

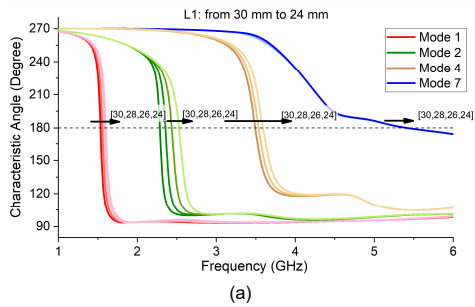

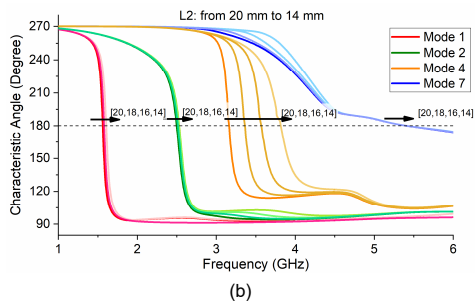

For this antenna, different bands are based on different modes, which exhibit different electric eigenfield distribution on the PCB, as shown in Fig. 12. For balancing the four modes, the CCE is located at the top-left of the PCB, while the feeding port is in the middle of the left side of the PCB, as shown in Fig. 11. The asymmetry of the CCE relative to the feeding port could help to improve the input impedance. Finally, after the excitation, we can also verify that the simulated real current distributions of the excited resonances are quite similar to the eigencurrent distributions of the above eigenmodes. The utilized modes are dominant in the four desired frequency bands.

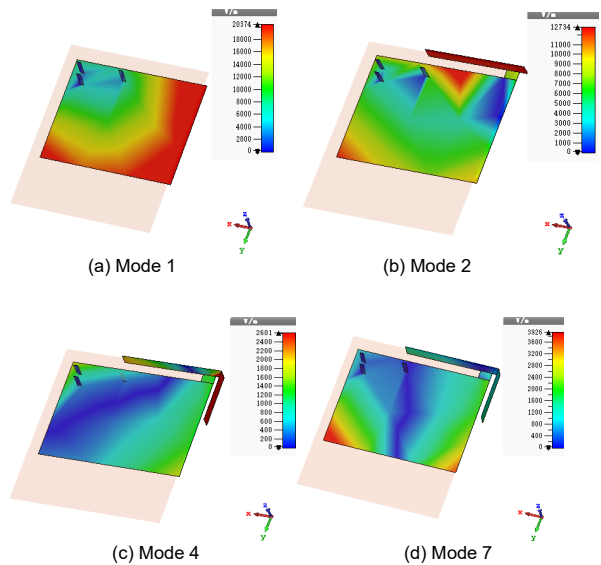

Fig. 12. Electric eigenfield distribution on the PCB of the utilized eigenmodes

TABLE I
DIMENSIONS OF THE PROPOSED ANTENNA

| Dimension | Length (mm) | Dimension | Length (mm) |
|---|---|---|---|
| $L_1$ | 28 | $hc$ | 4 |
| $L_2$ | 15 | $X_{CTX}$ | 20.5 |
| $L_c$ | 16 | $Y_{CTX}$ | 4 |
| $L_f$ | 3 | $X_{CTY}$ | 32.5 |
| $ps$ | 2 | $Y_{CTY}$ | 5 |
| $ss$ | 1 | $XX$ | 34 |
| $h$ | 4 | $YY$ | 28 |
| $hs$ | 3 | | |

## III. PROTOTYPE AND MEASURED RESULTS

After the fine-tuning of this model, the structure and detailed dimensions of the proposed antenna are shown in Fig. 11 and Table II. The size of the PCB is 34 mm × 28 mm. The size of the screen is 42 mm × 36 mm. The RO4003C 0.8-mm-thick laminates with a relative permittivity of 3.38 are used in the model for the screen and the PCB. The distance between the screen and the PCB is $h = 4$ mm. Because the total height of the smartwatch model is 10.7 mm, the PCB will lay in the middle of the smartwatch model, the same with the structure of *Apple*

*Watch Series 4* [48]. Thus, the distance between the PCB and the bottom of the smartwatch model is 5.1 mm.

$L_1 + L_2$ are the lengths of the stub. $L_c$ is the length of CCE. $L_f$ is the distance between the location of the feeding point and the lower end of the CCE. $ps$ is the slot width between the PCB and CCE. $ss$ is the slot width between the screen and the additional strip. $hs$ and $hc$ are the widths of the stub and CCE respectively. $(X_{CTX}, Y_{CTX})$ and $(X_{CTY}, Y_{CTY})$ are the coordinates of *CTX* and *CTY* respectively. The origin of the coordinate system is Point O, at the top right corner.

Then the prototype is fabricated. As in Fig. 11, the feeding port is on the left side of the PCB shown with a red arrow. A pigtail cable is mounted on the PCB from the left side to the right side. Then its SMA connector of the pigtail cable on the right side could be connected to a vector network analyzer (VNA) for further measurement. Fig. 13(a) & (b) show the top-right and top-left corner views. An impedance matching network is inserted to tune the antenna, near the antenna port on the left side of the PCB, as shown in Fig. 13(c). It could be designed with the assistance of ADS [49]. The matching network consists of a shunted 6.2 nH inductor, a series 0.3 nH inductor, and a shunted 1 pF capacitor. As mentioned in previous Section II(E), it is mainly for the tuning of 4.9 GHz mode. It could also improve the impedance matching of the four desired frequency bands.

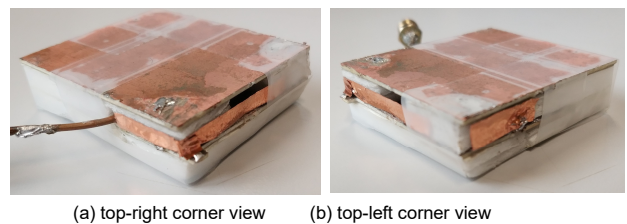

(a) top-right corner view　　(b) top-left corner view

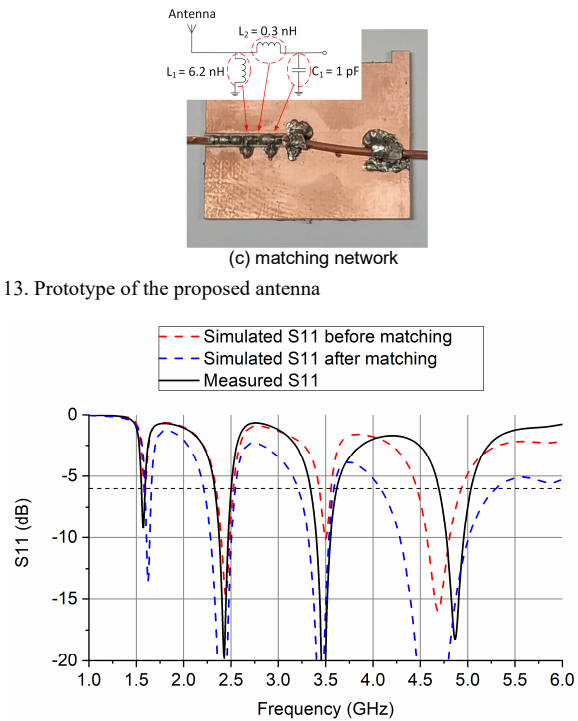

(c) matching network

Fig. 13. Prototype of the proposed antenna

Fig. 14. Simulated and measured reflection coefficient

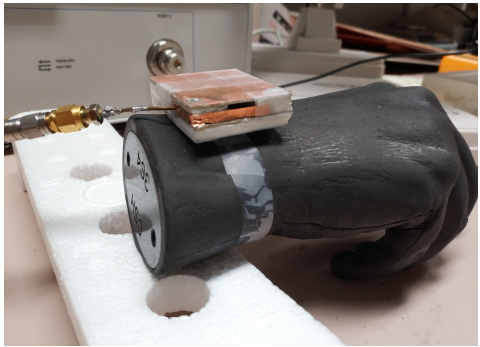

Fig. 15. Prototype on the hand phantom

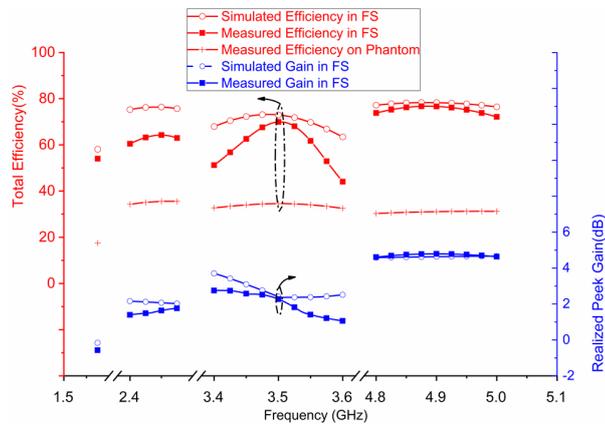

Fig. 16. Simulated and measured total efficiency and realized 3D peek gain in free space (FS) and on phantom

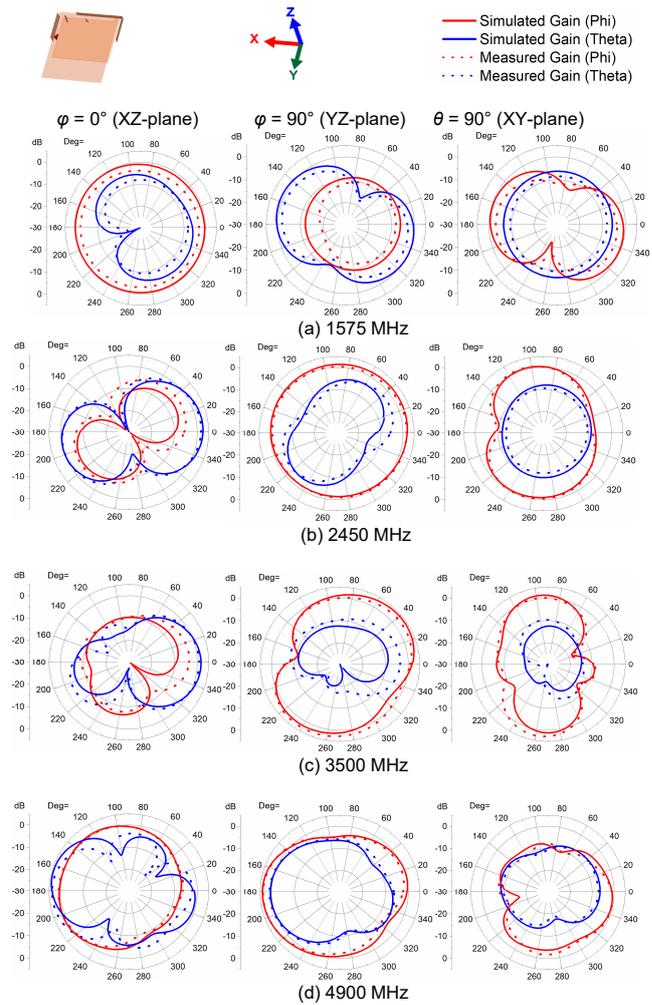

Fig. 17. Simulated and measured radiation patterns.

TABLE II
SUMMARY OF SMARTWATCH ANTENNAS

| | | Frequency bands | | | | | Full screen | Omnidirectivity | Antenna length (unit: mm) ↑ | × | Antenna width (unit: mm) | Free space efficiency* | Phantom efficiency* |
|---|---|---|---|---|---|---|---|---|---|---|---|---|---|
| | | GPS L1 | 2.45-GHz WLAN/BT | 3.5-GHz 5G | 4.9-GHz 5G | 5-GHz WLAN | | | | | | | |
| Single-band | [22] | | √ | | | | | √ | 17 | | 5 | 71% | 26% |
| | [29] | | √ | | | | | √ | 16 + 6 | | 6 & 2 | 50% | 46% |
| | [21] | | √ | | | | | √ | 37 | | 1 | 80% | 50% |
| | [24] | | √ | | | | | | 40 | | 5 | - | - |
| | [27] | | √ | | | | | √ | 48 | | 36 | 70% | - |
| | [23] | | √ | | | | | √ | 132 | | 7 | 75% | 42% |
| | [26] | | √ | | √ | | | √ | 132 | | 10 | 70% | 57% |
| | [25] | | √ | | | | | √ | 180 | | 5 | 70% | 26% |
| Multi-band | [31] | | √ | | √ | √ | √ | | 38 | | 32 | ≈25% | - |
| | [30] | | √ | √ | | | | √ | 138 | | 11.5 | ≈65% | - |
| | [19] | √ | √ | | | √ | | | 173 | | 1.5 | 62.9%** | - |
| | [18] | √ | √ | √ | √ | | | | 180 | | 5 | 73% | - |
| | [20] | | √ | √ | √ | | | | 180 | | 5 | 67%** | - |
| | **Proposed** | √ | √ | √ | √ | | √ | √ | **43+16** | | **3 & 4** | **67%** | **33%** |

*Measured results. Lower bound of the efficiency is adopted.    **Simulated results.

The simulated reflection coefficient before and after impedance matching, and the measured reflection coefficient in the free space of the prototype are shown in Fig. 14. The antenna could cover the four desired frequency bands with a reflection coefficient lower than -6 dB. Then, the prototype was measured in the SATIMO microwave chamber. The simulated and measured radiation patterns of the central frequencies of the four bands are shown in Fig. 17. At all the frequency bands, the antenna exhibits good omni-directional radiation patterns, which are suitable for smartwatches. This antenna is applied in mobile devices, whose rotation angles are always changing. Thus, even though the radiation directions and the cross-polarization levels are not stable at different frequencies, the reception of the dual-polarized base station antenna is not affected. The prototype is also measured when mounted on a standard hand phantom, as in Fig. 15. The simulated and measured total efficiency and realized peak gain in free space and on hand phantom are shown in Fig. 16. The measured total efficiency in free space is 67% in average, while it is 33% on phantom. The phenomenon that the efficiency on phantom is lower than that in free space is because of the hand phantom's absorption.

## IV. Conclusion

In this paper, the fundamental structure (including the full screen and the system PCB) of the smartwatch is analyzed as a whole by the theory of characteristic modes (TCM), then modified as the radiator for a multi-band antenna. This method helps to improve the performance of smartwatch antennas. In Table II, the proposed antenna is compared with smartwatch antennas in present literature in the aspects of frequency bands, compatibility to full screen, omnidirectivity of the radiation pattern, antenna size, and free-space and phantom efficiency. We can see that this method could design very small-size smartwatch antennas, add more frequency bands easily, radiate omni-directionally, and fit to the structures of present smartwatches.

From the above Fig. 8, we could also see that there are some unadopted characteristic modes which are available for more possible resonances. Moreover, we could also further modify the fundamental structure for more desired characteristic modes. These are all potential methods for developing more multi-band smartwatch antennas. In a word, this method is beneficial for the widespread connection in various communication protocols for smartwatches in IoT applications.